\documentclass[prl,aps,twocolumn,floatfix]{revtex4}
\usepackage{epsf}
\usepackage{epsfig}
\usepackage{graphics}
\begin{document}

\title{
Pairing Fluctuations and Anomalous Transport Above the BCS-BEC Crossover\\
in the Two Dimensional Attractive Hubbard Model
}

\author{Sabyasachi Tarat and Pinaki Majumdar}

\affiliation{Harish-Chandra  Research Institute,
 Chhatnag Road, Jhusi, Allahabad 211019, India}

\date{5 May 2011}

\begin{abstract}
A Fermi liquid with weak attractive interaction undergoes a BCS transition 
to a superconductor with reducing temperature. With increasing interaction 
strength, the thermal transition is progressively modified as the high 
temperature `metallic' phase develops a pseudogap due to pairing fluctuations 
and the resistivity above $T_c$ shows insulating behaviour. The crossover to 
insulating character occurs much before the system can be considered to be in 
the BEC regime of preformed fermion pairs. We use a new Monte Carlo tool to 
map out the BCS-BEC crossover in the attractive Hubbard model on large two 
dimensional lattices and explicitly compute the resistivity to demonstrate how 
the metal to superconductor (MS) thermal transition at weak coupling crosses 
over to an insulator to superconductor (IS) transition at intermediate coupling. 
Our high resolution access to the single particle and optical spectrum at finite
temperature allows us to completely describe the transport crossover in this 
longstanding problem.
\end{abstract}

\maketitle

The BCS-BEC crossover in attractive fermion systems has been a
topic of interest \cite{bcs-bec-rev} for several decades.
With increasing interaction, the ground state of a weak coupling 
`BCS superconductor', with pair size $\xi$ 
much larger than the interparticle separation $\sim k_F^{-1}$
(where $k_F$ is the Fermi wavevector)
evolves smoothly \cite{eagles,leggett,noz,micnas,rand-rev}
into a `Bose-Einstein condensate' (BEC)
of preformed fermion pairs with $\xi \lesssim  k_F^{-1}$.
The `high temperature' normal state changes from a 
conventional Fermi liquid at weak coupling to a gapped phase 
at strong coupling. While the pairing gap increases with
coupling strength,  
the superconducting $T_c$ in lattice models reaches a
maximum at intermediate coupling and falls thereafter.
A striking consequence of the separation of
pairing and superconducting scales is the emergence of
a (pseudo)gapped normal phase, 
with preformed fermion pairs but no
superconductivity  due to strong phase fluctuations.

The early work of Leggett \cite{leggett}
and Nozieres  and Schmitt-Rink \cite{noz} provided the intuitive
basis for understanding this problem. It has since been 
followed up by extensive quantum Monte Carlo (QMC) work 
\cite{qmc-scal1,qmc-moreo1,qmc-moreo2,sc-spingap,sc-nfl},
powerful semi-analytic schemes 
\cite{t-matrix1,t-matrix2,t-matrix3}, and most recently
dynamical mean field theory (DMFT) \cite{dmft1,dmft2}.
The efforts have established the presence of a
pseudogap \cite{qmc-moreo2}  
in the single particle spectrum beyond moderate coupling
and temperature $T > T_c$, and also a gap in the spin 
excitation spectrum \cite{sc-spingap}.
While these indicate a breakdown of the Fermi liquid picture, 
the crucial {\it transport properties}
 in the normal state remain obscure. 

For example, given the `gapped' normal state at 
strong coupling, is it insulating? If so, down 
to what coupling does this extend? 
There seem to be various options: (i)~the insulating
state arises only 
when the single particle density of states (DOS) at
$T>T_c$ has a `hard gap', or
(ii)~metallic conduction survives even in the hard gap 
state due to transport by `bosonic' carriers, or (iii)~the
normal state becomes insulating, due to
strong `pairing disorder', {\it even without} a hard gap 
in the spectrum.
QMC calculations, which set the
benchmark in the field, unfortunately do not have access
to real frequency information or the system size
that can resolve this issue.

We use a new Monte Carlo method, involving a Hubbard-Stratonovich
(HS) decomposition \cite{solms}
of the attractive Hubbard model in terms of pairing
fields, in two dimensions (2D). 
We ignore the time fluctuations of the (bosonic) 
pairing field treating it as classical \cite{dubi}, 
but fully retain the spatial fluctuations of its  
amplitude and phase. 
Our principal results are the following:
(1).~We benchmark the $T_c$ obtained by our method with
the most recent QMC results, demonstrating the
accuracy of our method at temperatures of interest. (2).~We
obtain the resistivity $\rho(T)$ over the entire coupling
range, and observe that the high temperature phase goes insulating
at $U/t \gtrsim 3$, in the nominally `BCS regime' and
much before the occurence of preformed pairs. 
(3).~The single particle DOS reproduces features 
previously inferred 
from QMC, but the optical spectrum behaves very differently
from the single particle DOS, `filling up' 
at a much lower temperature at strong coupling.


We study the attractive Hubbard model in 2D.
\begin{equation}
H = - t\sum_{\langle ij \rangle \sigma} c_{i \sigma}^{\dagger} c_{j \sigma} 
- \mu \sum_{i \sigma} n_{i \sigma} 
- \vert U \vert \sum_{i} n_{i \uparrow} n_{i \downarrow}
\end{equation}
$t$ denotes the nearest neighbour tunneling amplitude, 
$U >0$ is the strength of onsite attraction, and $\mu$ the
chemical potential. We will study the range $ 1 \le U/t \le 10$,
going across the BCS-BEC crossover, and set $\mu$ so that
the particle density remains at $n \approx 0.9$.

The model is known to have
a superconducting (S) ground state for all $n \neq 1$, while at
$n=1$ there is the coexistence of superconducting and  density wave (DW) 
correlations in the ground state. 
For $ n \neq 1$ the ground state evolves from a
BCS state at $U/t \ll 1$ to a 
BEC of `molecular pairs' at $U/t \gg 1$.  
The pairing amplitude and gap at $T=0$ 
can be reasonably accessed within mean field theory
or simple variational wavefunctions.

Mean field theory, however, assumes that the
electrons are subject to a {\it spatially uniform} 
self-consistent pairing amplitude $  \langle \langle 
c^{\dagger}_{i \uparrow} c^{\dagger}_{i \downarrow} \rangle \rangle$.
At small $U/t$ this vanishes when $k_B T \sim 
te^{-t/U}$, but at large 
$U/t$ it vanishes only when  $k_B T \sim U$. The 
actual $T_c$ at large $U$ is controlled by phase correlation 
of the local order parameter, rather than finite
pairing amplitude, 
and occurs at $k_B T_c \sim f(n)t^2/U$, where $f(n)$ is a 
function of the density. 
The wide temperature 
window, between the `pair formation' scale $k_B T_f
\sim U$ and $k_B T_c$ corresponds to equilibrium between
unpaired fermions and hardcore bosons (paired fermions). 
QMC calculations have suggested a single particle pseudogap
and a gap in the spin  (NMR) excitation spectrum for $U/t \gtrsim
4$ and $T > T_c$.  None of the calculations, however, seem 
to have addressed the simplest measurable property, {\it i.e},
normal state charge transport. This is probably due to
`analytic continuation' problems in QMC data or
the severe finite size effects in exact diagonalisation based
schemes.

We use a strategy used earlier to access the superconductor to
insulator transition in the disordered attractive Hubbard model
\cite{dubi}, and models of $d-wave$ pairing \cite{dag}, 
augmented now by a Monte Carlo technique that
allows access to system size upto $\sim 40 \times 40$, 
much larger
than the coherence length for the chosen $U \ge 2$.
We decouple the Hubbard term in the pairing channel by
using the HS transformation and treat the HS field 
in the static approximation.
This gives qualitatively correct answers at all $n \neq 1$, and
surprisingly accurate $T_c$ values when compared to full QMC
\cite{vnand}.

The static HS approach leads to the effective model:
\begin{equation}
H_{eff}  =  H_0  +
\sum_{i} ( \Delta_{i} c_{i \uparrow}^{ \dagger}
c_{i \downarrow}^{ \dagger } +
\Delta_{i}^{ \star } c_{i \downarrow } c_{ i \uparrow }) +
\sum_{i} \frac{\vert \Delta_{i} \vert^{2}}{U}
\end{equation}
where $H_0 =
- t\sum_{\langle ij \rangle \sigma} c_{i \sigma}^{\dagger} c_{j \sigma}
- \mu \sum_{i \sigma} n_{i \sigma}
$ and $\Delta_i = \vert \Delta_i \vert e^{i \theta_i}$ is a
complex scalar
{\it classical} field. This model allows fluctuations in both
the amplitude and phase of the HS variable,  and the
fermions propagate  typically in an inhomogeneous background
defined by $\Delta_i$.

\begin{figure}[b]
\centerline{
\includegraphics[width=4.1cm,height=3.5cm,angle=0]{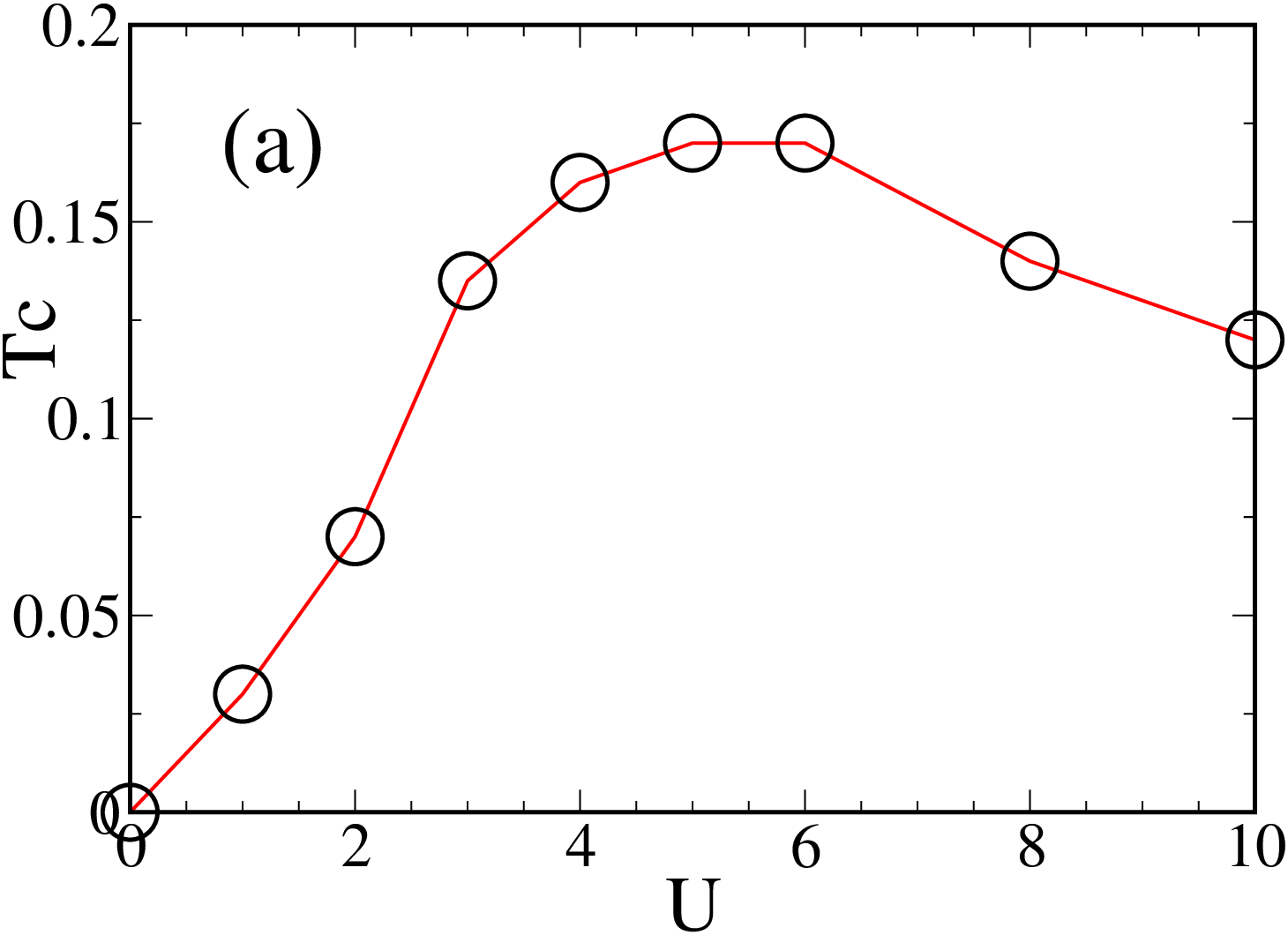}
\includegraphics[width=4.1cm,height=3.4cm,angle=0]{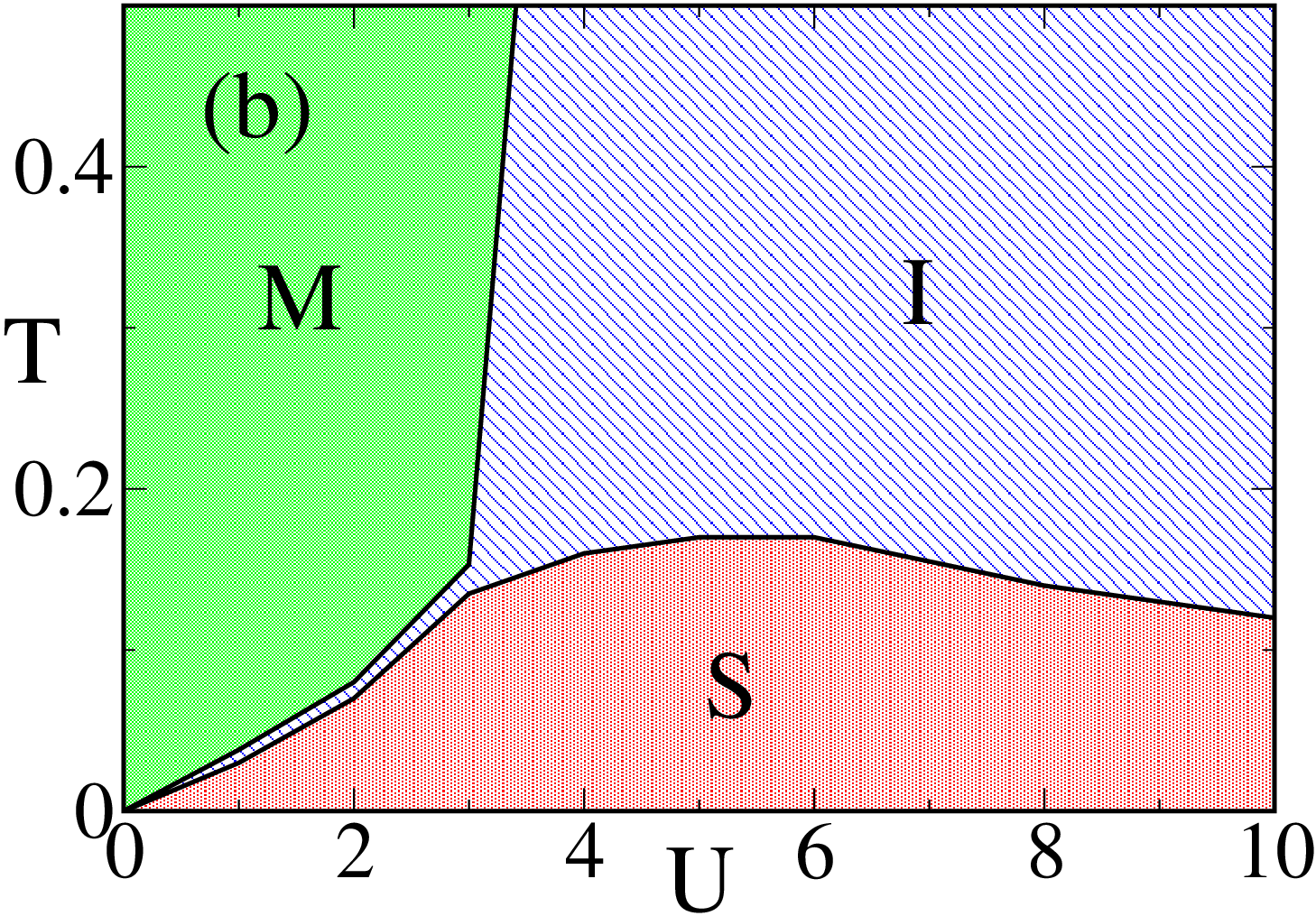}
}
\vspace{.6cm}
\centerline{
\includegraphics[width=4.1cm,height=3.5cm,angle=0]{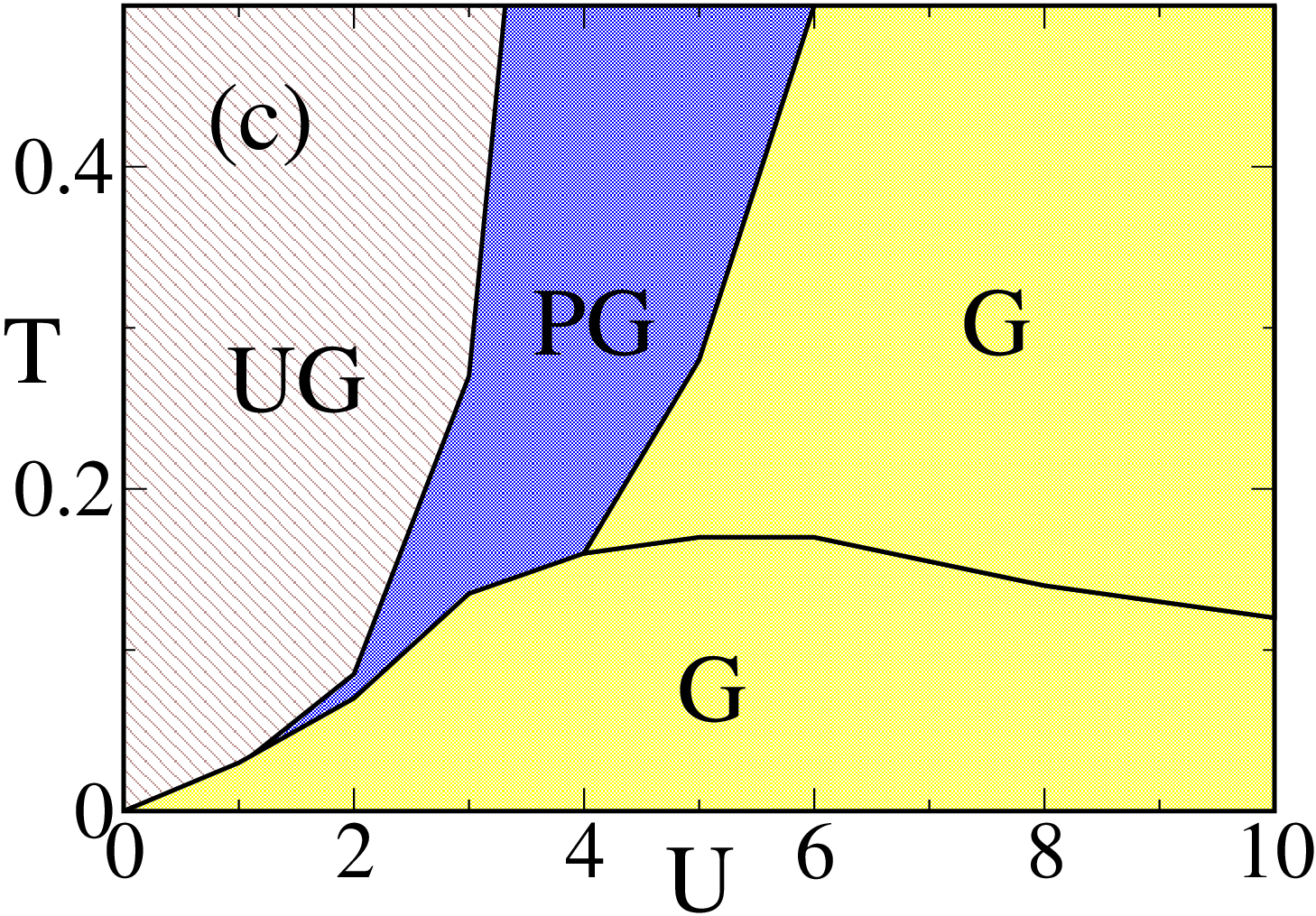}
\includegraphics[width=4.1cm,height=3.5cm,angle=0]{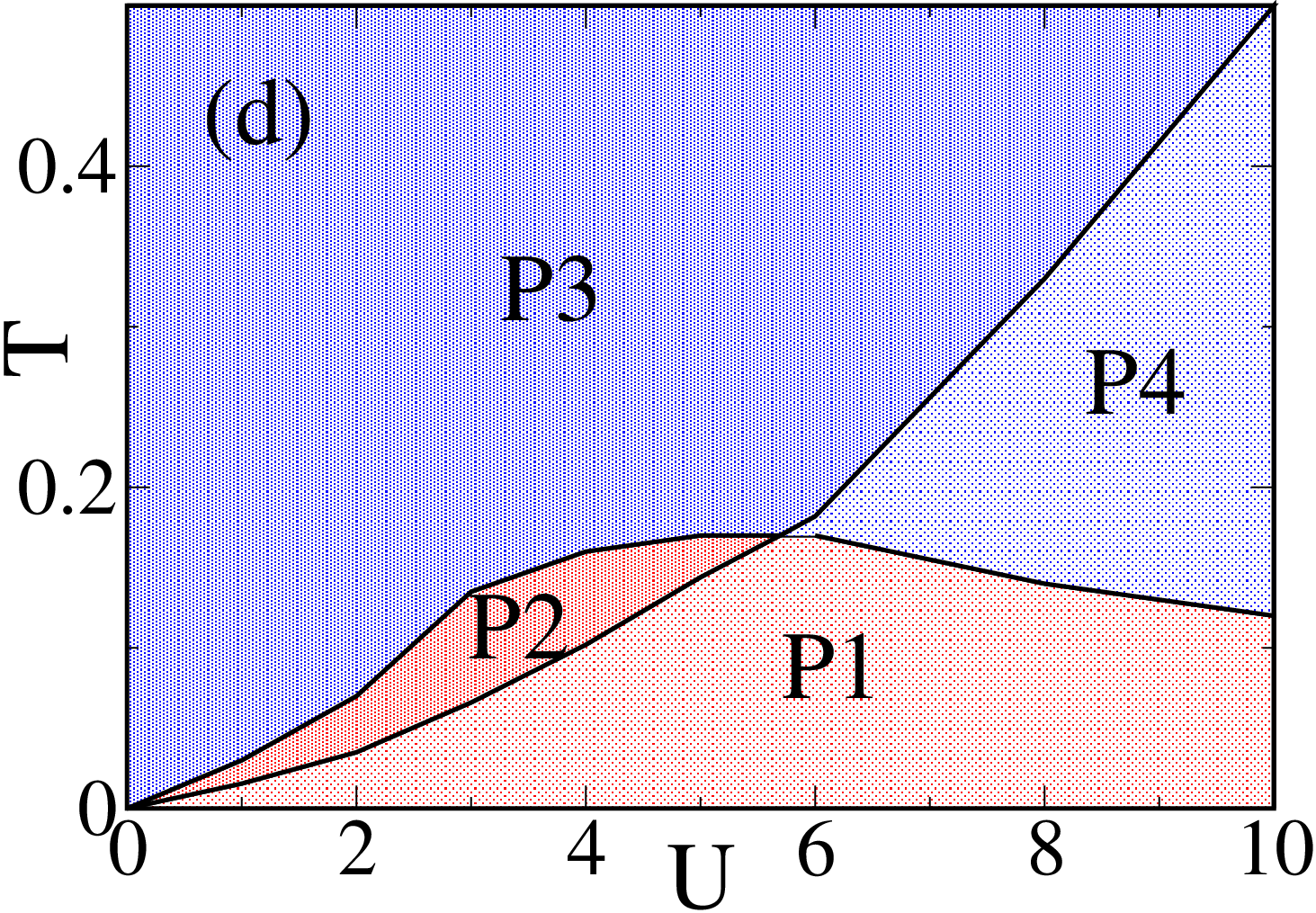}
}
\caption{Colour online: (a).~The transition temperature $T_c(U)$.
(b).~Superconductor-metal-insulator (S-M-I) character. The superconductor
has $\rho_{dc}=0$, the M and I have $\rho_{dc}$ finite, with $d\rho_{dc}/dT
>0$ for $M$ and $d\rho_{dc}/dT <0$ for $I$. (c).~`Phases' according
to single particle DOS near Fermi level: 
UG is ungapped (Fermi liquid), PG is pseudo-gapped, and G is gapped.
By this criteria the superconductor and $T > T_c$ insulator
are similar. (d).~Classification according to low frequency 
optical spectrum, {\it i.e}, the presence of the superfluid
$\delta$ function, and a continuous weight extending to
$\omega =0$. 
Both P1 and P2 are in the superfluid region but also have low frequency
optical spectral weight (except at  $T=0$). The distinction between P1 and
P2 is quantitative, P2 has much larger low $\omega$ weight.
P3 and P4 are both `normal' {\it i.e}, without a superfluid peak,
and also have low $\omega$ weight. Again, the P3, P4 distinction is
quantitative, the low $\omega$ weight is exponentially small in
P4.
}
\end{figure}

\begin{figure}[t]
\centerline{
\includegraphics[width=7.0cm,height=6.0cm,angle=0]{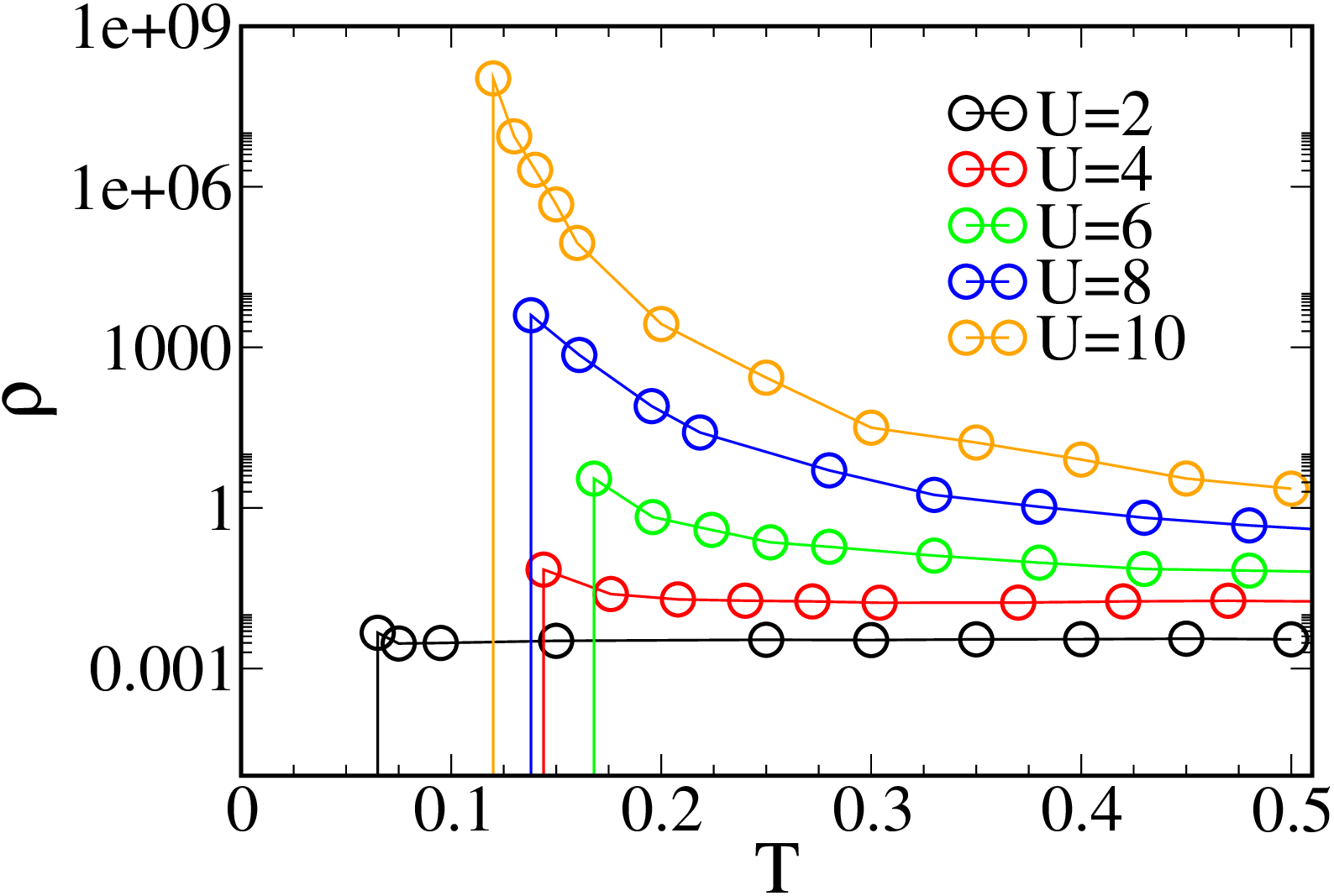}
}
\caption{Colour online: The resistivity $\rho(T)$ for various
interaction strengths. There is a region with $d\rho/dT <0$ above
$T_c$ at all $U$, including the weak $U$ systems. Note that the
$y$ axis spans 12 orders of magnitude.
}
\end{figure}

To obtain the ground state, and in
general configurations $\{ \vert \Delta_i \vert,
\theta_i\}$ that
follow the distribution $P\{ \vert \Delta_i \vert,\theta_i\}
\propto e^{-\beta H_{eff}}$, we use the Metropolis algorithm to
update the $\vert \Delta \vert$ and $\theta$ variables.
This involves solution of the Bogoliubov-de Gennes (BdG)
equation \cite{bdg} for each attempted update. For equilibriation 
we use a `traveling cluster'
algorithm \cite{tca}, diagonalising the BdG equation on a $8 \times 8$
cluster around the update site.
Global properties  like pairing field correlation, DOS, 
{\it etc}, are computed via solution of the BdG equation
on the full system.
All results in this paper are for system size $32 \times 32$.

If $\epsilon^{\alpha}_n$ are the BdG eigenvalues in some equilibrium
configuration $\alpha$, the quasiparticle DOS is computed
as $N(\omega) = \langle \sum_n \delta(\omega - \epsilon^{\alpha}_n)
\rangle $, where the angular brackets indicate averaging over
$\alpha$. The quasiparticle (QP)
 gap is the minimum of $\epsilon^{\alpha}_n$ over
all $\alpha,n$ at a given $T$.
Similarly, the 
optical conductivity in an equilibrium background is formally
$ \sigma(\omega) = -\omega^{-1} Im(\Lambda_{xx}(q=0,\omega)) $
where the current-current correlation function is defined by
$$
\Lambda_{xx}(q=0, \omega) = 
\frac{1}{ \cal{Z}} \sum_{n,m} | \langle n | j_{xx} | m \rangle |^{2} 
\frac{e^{- \beta E_{n}}- e^{- \beta E_{m}}}{\omega + E_{n}- E_{m}+ i \delta}
$$
The $\vert n \rangle$, $\vert m \rangle$ 
are multiparticle states of the system. 
We have suppressed the $\alpha$ labels above.
We will discuss the simplification of this expression 
elsewhere. For the continuous part of $\sigma(\omega)$,
{\it i.e}, excluding the superfluid response, it leads to:
\begin{eqnarray}
\sigma(\omega) & =& 
\sum_{a,b} F_1(a,b)  
\frac{(n(\epsilon_{a}) + n(\epsilon_{b}) - 1 )}
{\epsilon_{a} + \epsilon_{b}} 
\delta(\omega - \epsilon_{a}- \epsilon_{b}) \cr 
&&~+ \sum_{a,b} F_2(a,b)  
\frac{(n(\epsilon_a ) - n(\epsilon_{b}) )}{\epsilon_{a} - \epsilon_{b}} 
\delta(\omega - \epsilon_{b} + \epsilon_{a}) 
\nonumber
\end{eqnarray}
where, now, the $\epsilon_{\alpha}, \epsilon_{\beta} 
> 0$, {\it etc}, are {\it single particle
eigenvalues} of the BdG equations, the $n(\epsilon_a)$, {\it etc.},
are Fermi functions, and the $F$'s are
current matrix elements 
computed from the BdG eigenfunctions. 
We will call the two contributions above, $\sigma_1(\omega)$ and
$\sigma_2(\omega)$.
The averaging of $\sigma(\omega)$ over
equilibrium configurations of $\Delta_i$ at a given $T$ is implied.
The dc resistivity  $\rho =  {\omega_0}^{-1} 
\int_0^{\omega_0}  \sigma(\omega) d \omega$, where $\omega_0 
\sim 0.1t$.


{\it Phase diagram:}
Fig.1.(a)  shows our result on $T_c(U)$, at $n=0.9$.
We compute the thermally averaged 
pairing field correlation
$S({\bf q}) = {1 \over N^2} \langle \sum_{ij} 
\vert \Delta_i \vert \vert \Delta_j \vert
cos(\theta_i - \theta_j) e^{i {\bf q}.{(\bf r}_i -{\bf r}_j)}
\rangle $ at ${\bf q} = \{0,0\}$. This is like the `ferromagnetic' correlation
between the $\Delta_i$, treating them as two dimensional moments. If
the ${\bf q} = \{0,0\}$ component, $S(0)$, is ${\cal O}(1)$ it implies
that the pairing field has a non-zero spatial average and would in
turn induce long range order in the thermal and quantum 
averaged correlation $ M_{ij} =
\langle \langle c^{\dagger}_{i\uparrow} c^{\dagger}_{i\downarrow}
c_{j\downarrow} c_{j\uparrow} \rangle \rangle$. 
We locate the superconducting
transition from the rise in $S(0,T)$ as the system is cooled. 
The results are not reliable
below $U/t \lesssim 1$, since $\xi$ 
becomes comparable to our system size, but 
compare very well with available QMC data for $U/t \gtrsim 2$.

Fig.1.(b) shows the basic classification of the $U-T$ space
in terms of metal (M), insulator (I) and superconductor (S),
the M and I regions determined from the slope $d \rho/dT$
(see Fig.2). The remarkable feature is that for $U/t
\gtrsim 3$ the system is insulating, way before one can 
invoke a `hard' gap in the spectrum due to `preformed pairs'.
Fig.1.(c) highlights the low energy behaviour in the
quasiparticle density of states (DOS). The superconducting phase has
a gap for all $U$ and $ T < T_c$. So does the large $U$ normal
state, as indicated. The low $U$ system has a band like DOS for
$T > T_c$ and the most intriguing behaviour occurs for
$2 \lesssim U/t \lesssim 5$, where the $T > T_c$ phase has
a `dip' in the low energy DOS, {\it i.e}, a pseudogap.

\begin{figure}[b]
\centerline{
\includegraphics[width=4.0cm,height=3.0cm,angle=0]{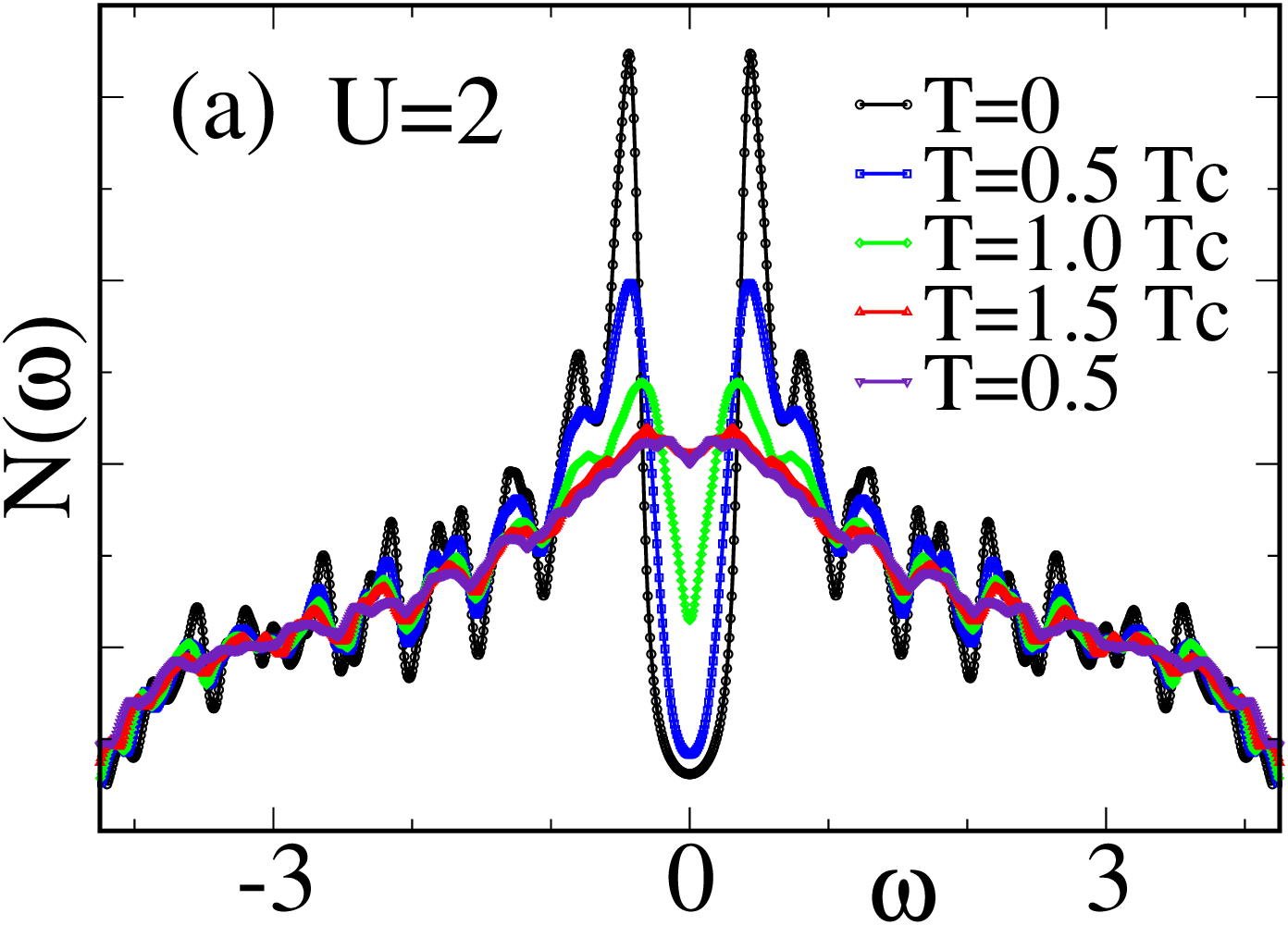}
\hspace{.3cm}
\includegraphics[width=4.0cm,height=3.0cm,angle=0]{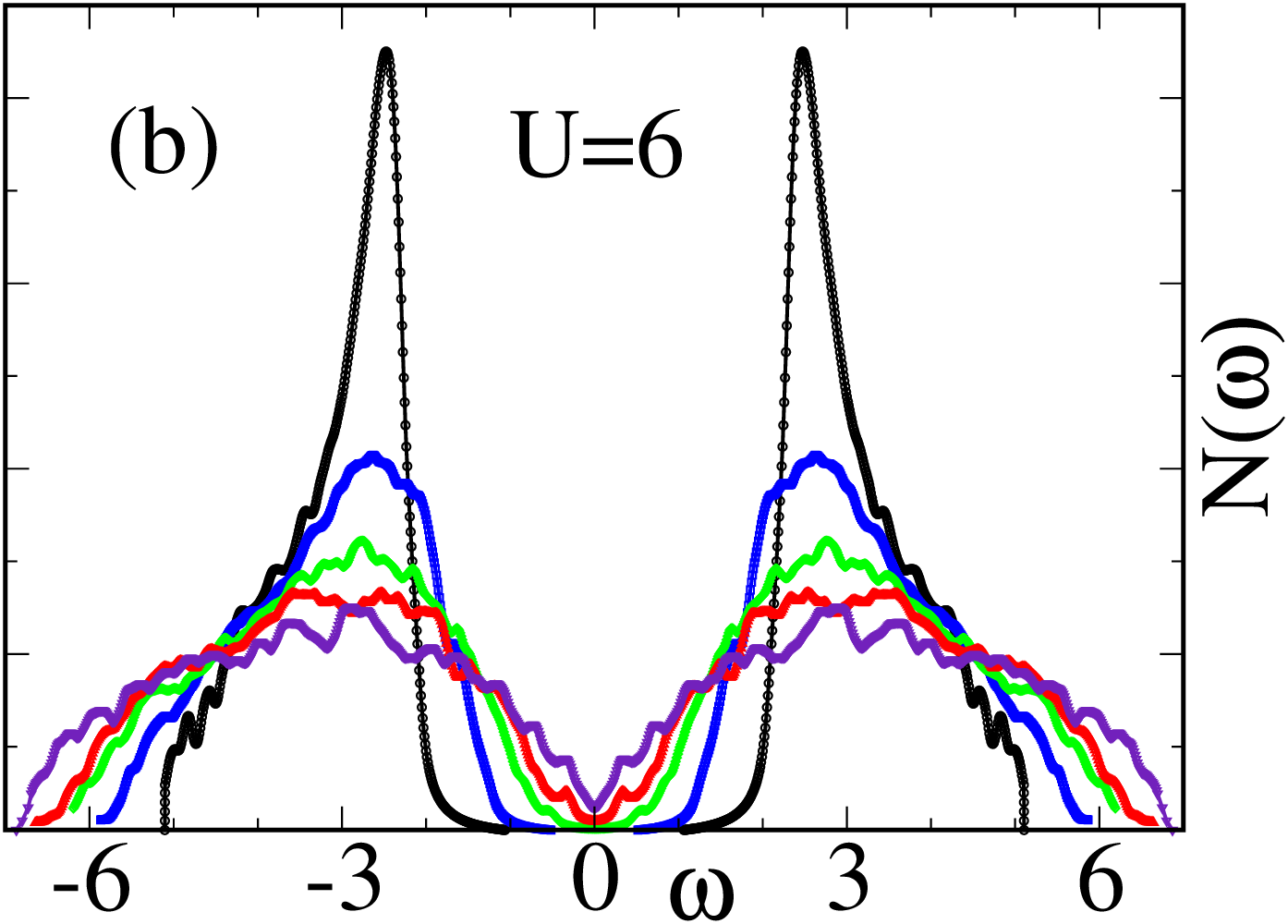}
}
\vspace{.5cm}
\centerline{
\includegraphics[width=4.0cm,height=3.0cm,angle=0]{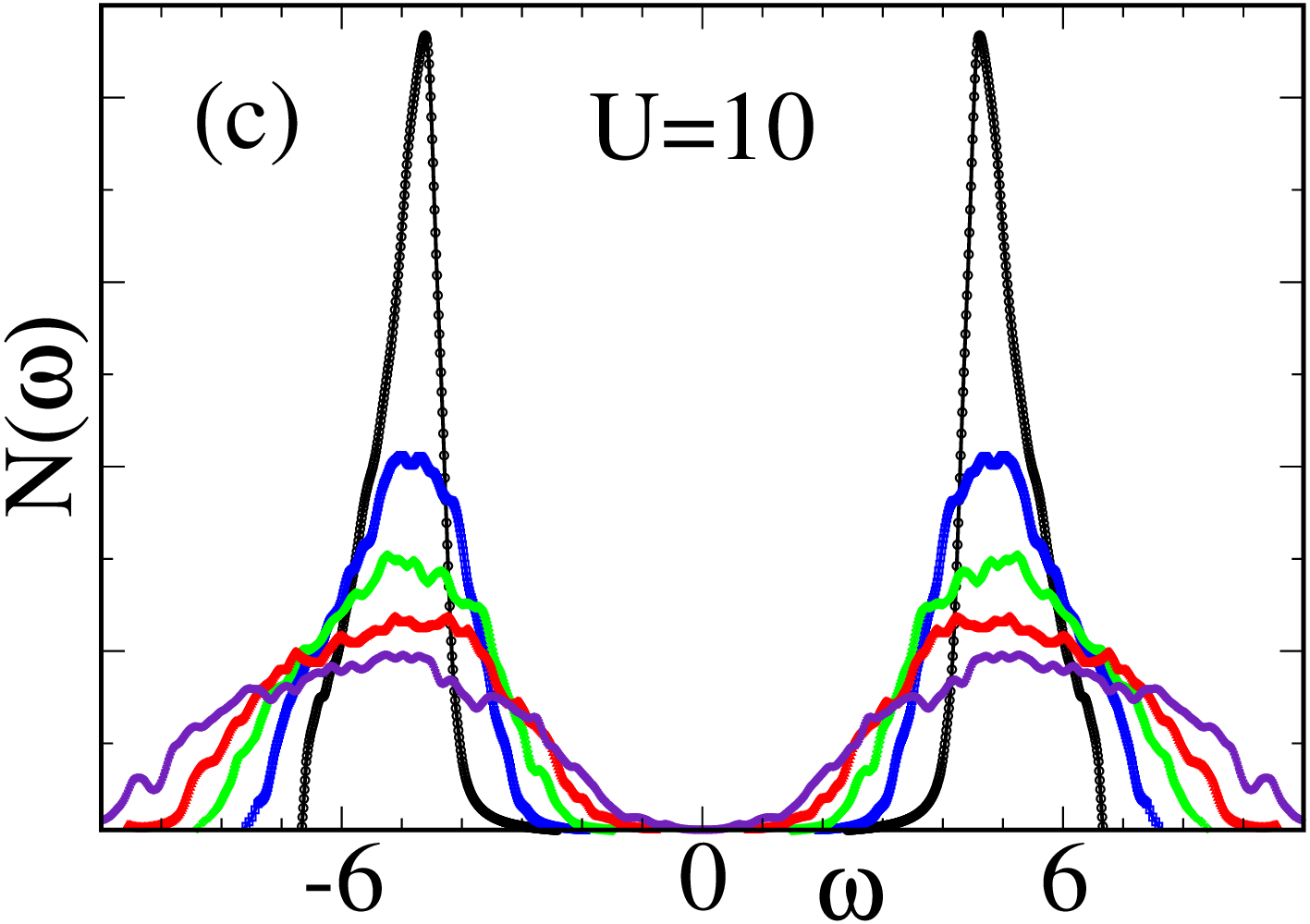}
\hspace{.3cm}
\includegraphics[width=4.2cm,height=3.0cm,angle=0]{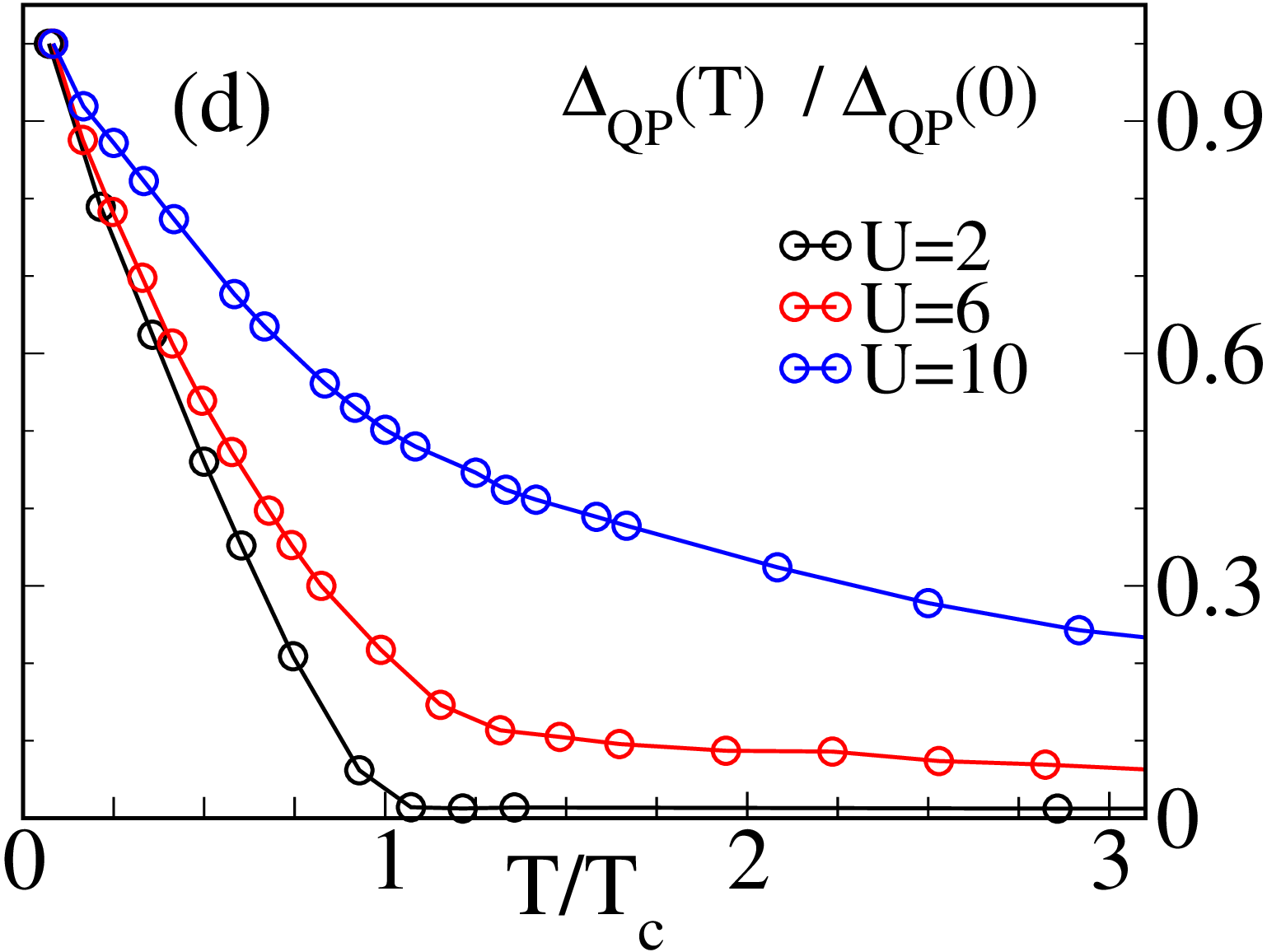}
}
\caption{Colour online: Temperature dependence of the
QP DOS, $N(\omega)$ at different couplings.
Panels (a)-(c) have the same legends.
(a).~Weak coupling, $U/t=2$, (b).~intermediate coupling, $U/t=6$,
and (c).~strong coupling, $U/t=10$. The oscillations in the DOS
in panel (a) are finite size artifacts (even on a $32 \times 32$
lattice). At $U/t=2$ the gap essentially vanishes at $T \sim T_c$,
while at $U/t=6$ a small `hard gap' persists to  $T_c$ and
above, although lorentzian broadening gives the impression of
a pseudogap at the highest $T$.
For $U/t=10$ a `hard gap' persists to
$T \sim 0.5$ although with a clear
reduction with increasing temperature.
Panel (d).~shows the variation in the gap, normalised
by its $T=0$ value,
for the three couplings. The gap is inferred
directly from the eigenvalue spectrum while the DOS has a
lorentzian broadening.
}
\end{figure}

Fig.1.(d) shows how the optical spectrum varies for changing
$U$ and $T$. 
As the expression for $\sigma(\omega)$ reveals, except at 
$T=0$, there is in principle always low frequency spectral
weight in the optical conductivity at any $U$ and $T$.
This arises from thermally excited quasiparticles, and is 
exponentially 
small when $\Delta_{QP}/k_BT \gg 1$, {\it i.e}, broadly
in the lower right part of the $U/T$ plane (the P1, P4
regions). 
Following the same argument, P1 and P3 have significant 
low frequency weight. P1 and P2 of course also have
a superfluid $\delta$ function feature which P3 and P4 
do not have.

{\it Resistivity:} Let us shift to the resistivity, Fig.2. 
From the expression for $\sigma(\omega)$ it is obvious
that as long as there is a gap in the QP spectrum, the
$\sigma_1$ term cannot contribute to the dc conductivity
(it has a lower cutoff $2 \Delta_{QP}$). We tried a crude 
model to analyse the $\rho(T)$ result. We assumed 
$\Delta_i$ configurations such that $\vert \Delta_i \vert$
is {\it same at every site} and is set to the $T=0$ mean 
field value appropriate to the $\mu$ and $U$. The phases
$\theta_i$ are assumed to be completely random and
uncorrelated between sites. Solving the BdG equations 
that results from these configurations, and introducing
$T$ only as in the Fermi factors, leads to a result that
is remarkably similar to Fig.2 for $T \gtrsim 0.3$. This
suggests that the detailed $T$ dependent distributions of
$\vert \Delta_i \vert $ and the spatial correlations in $\theta_i$ 
are not essential for a first understanding. The $\rho(T)$
for $U \gtrsim 6$ is mainly controlled by QP activation
across the $T > T_c$ gap, while for $U \lesssim
6$ we observed a pseudogap in the spectrum generated by 
the toy $\{\Delta_i\}$ and the $\rho(T)$ behaviour is
affected mainly by the scattering effect due to the random
$\theta$.  The M-I transition, with growing $U$ at $T > T_c$,
is mainly due to the scattering induced by pairing (angular)
disorder, rather than the opening of a clean gap in the 
QP spectrum.

\begin{figure}[t]
\centerline{
\includegraphics[width=5.4cm,height=3.5cm,angle=0]{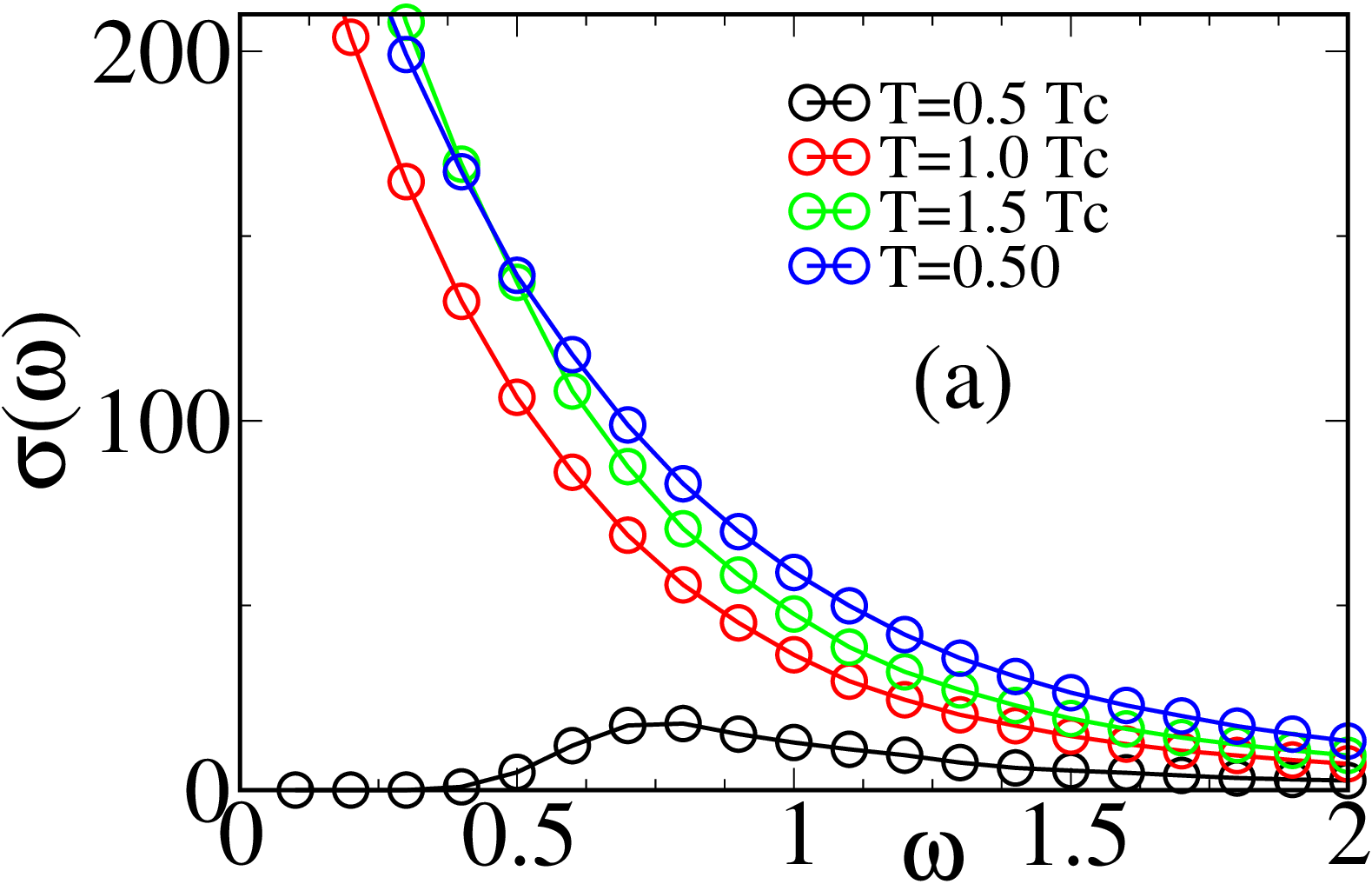}
}
\vspace{.6cm}
\centerline{
\includegraphics[width=5.4cm,height=3.5cm,angle=0]{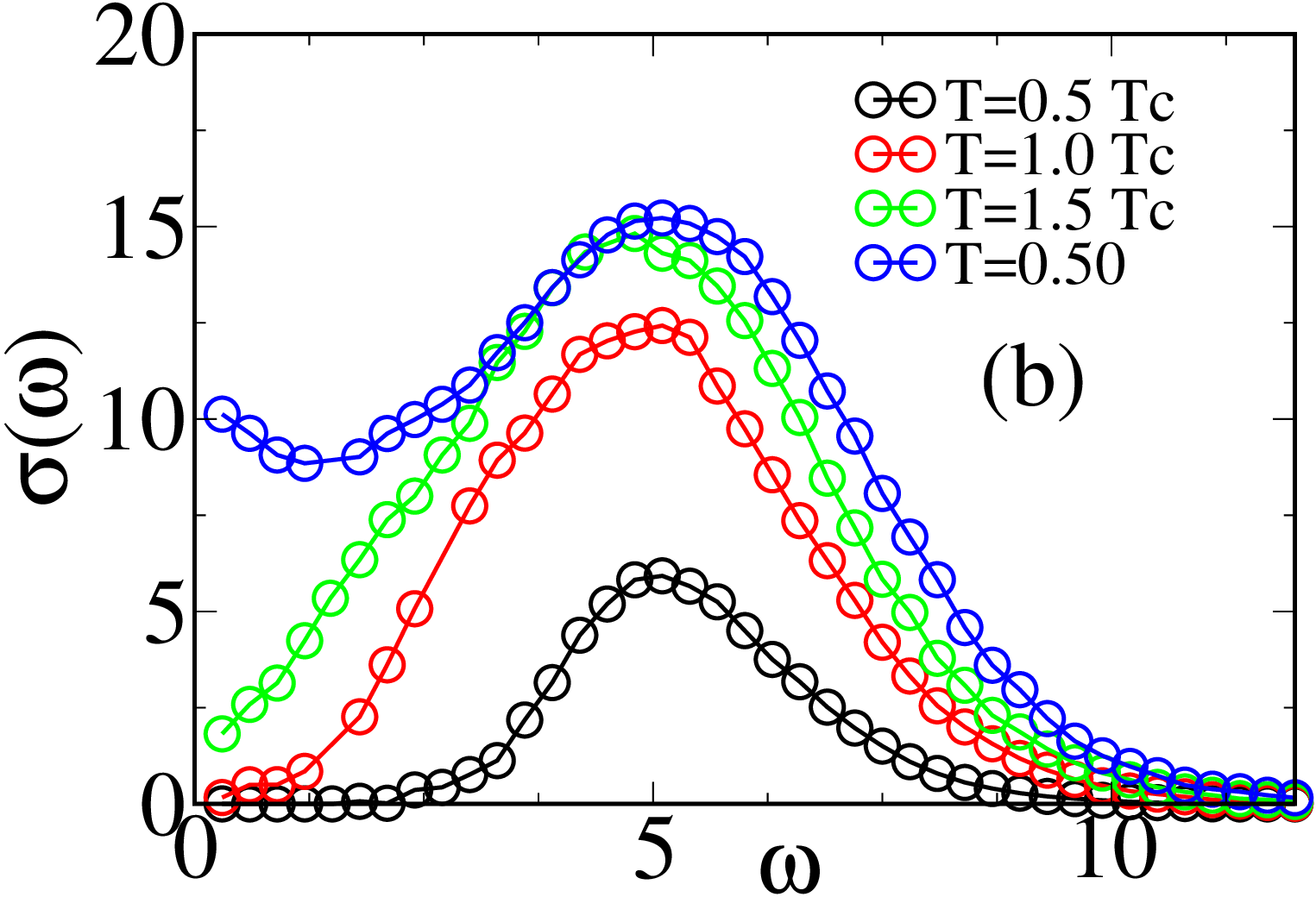}
}
\vspace{.6cm}
\centerline{
\includegraphics[width=5.4cm,height=3.5cm,angle=0]{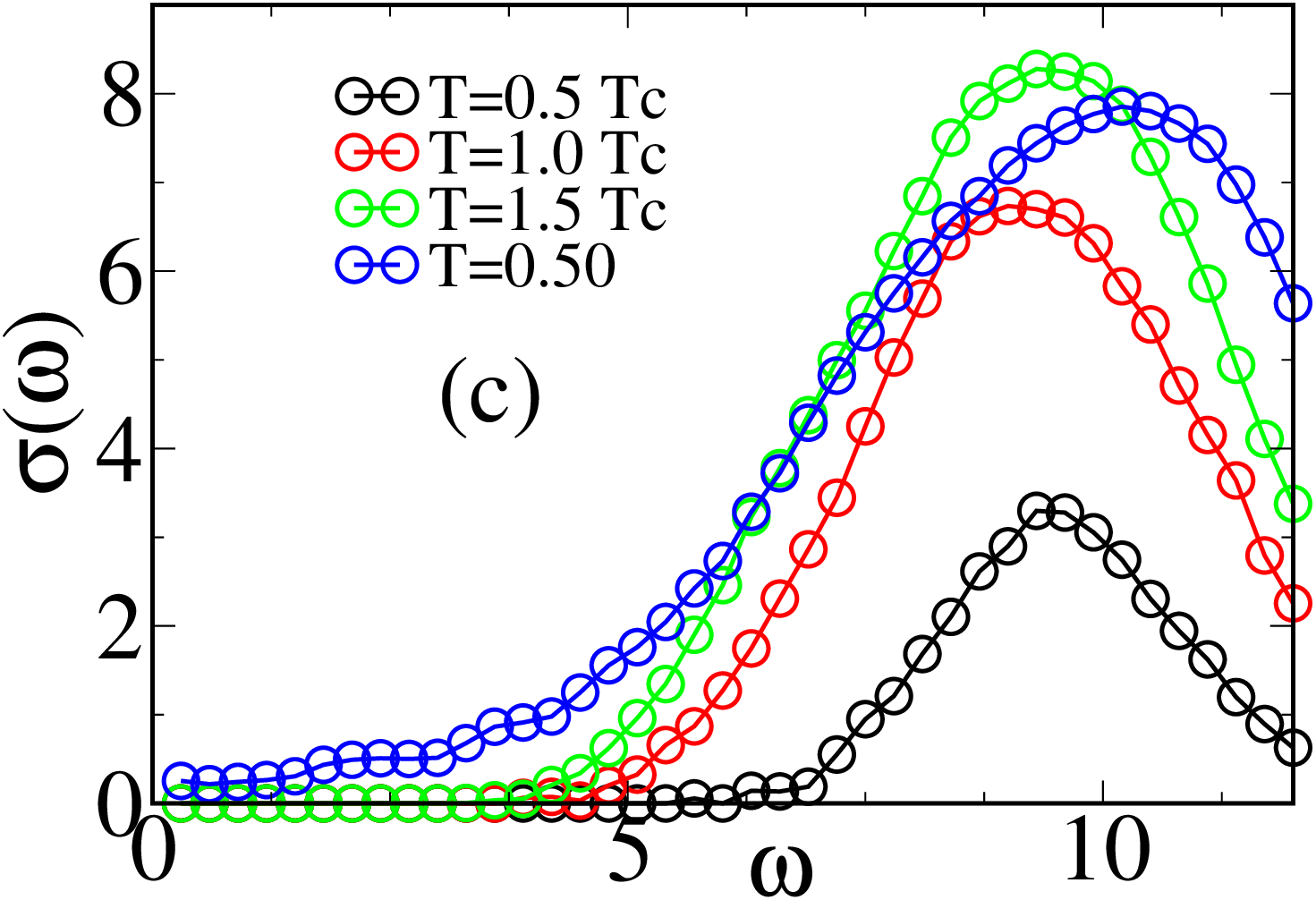}
}
\caption{Colour online: Temperature dependence of 
the optical conductivity, $\sigma(\omega)$
in different coupling regimes. 
(a).~For `weak' coupling: $U/t=2$, the low $\omega$ 
region fills up rapidly and $\sigma(\omega)$ is almost
$T$ independent for $T >T_c$. (b).~For intermediate coupling, $U/t=6$,
the suppressed low $\omega$ weight 
persists to $T \gtrsim T_c$, and the
overall weight in the 
$T > T_c$ spectra is moderately $T$ dependent. 
(c).~For $U/t=10$ the
suppression of low $\omega$ weight 
persists to $T \sim 0.5$ and the integrated weight seems
to increase slowly for $T > T_c$.
}
\end{figure}

{\it Density of states:} The DOS has been investigated earlier too,
and we confirm the known trends albeit with much higher resolution.
Panels (a)-(c) in Fig.3 show the DOS at $U=2,~6,~10$, 
respectively to the left of peak, peak $T_c$, and
right of peak, on the $T_c(U)$ curve. At $T=0$ all plots have the
usual coherence peak at the gap edges. These diminish rapidly with
$T$ and the gap begins to fill up. The fate of the gap is shown
in Fig.3.(d), where $U=6$ and $10$ have a gap for $T \gg T_c$
(and the $U=10$ gap is strongly $T$ dependent) while the $U=2$
gap closes at $T \sim T_c$. This confirms that $U \sim 3$, where
the M-I crossover occurs, will not have a clean gap for $T > T_c$.

{\it Optics:} Fig.4 shows the optical conductivity at $U=2,~6,~10$.
We have not shown the $\delta$ function at $T < T_c$ for clarity
(at $T=0$, within this calculation, the entire weight would be in
the $\delta$ function). The response at $U=2$ does not have any
finite $\omega$ peak
but $U=6$ and $U=10$
have a $2\Delta_{QP}$ peak, arising from $\sigma_1(\omega)$,
 that show up at $T \sim 0.5T_c$, and grow with increasing $T$. 
The filling up of the low $\omega$ part is however due to
thermally excited QP's contributing via $\sigma_2(\omega)$.
For $T >T_c$ the optical spectral weight weakly $T$ dependent.

In conclusion, we have studied the attractive 2D Hubbard model 
via a new Monte Carlo technique on large lattices. Our results
on $T_c$ and density of states correspond to available QMC 
results, but our access to real frequency conductivity data
allows the first determination of the metal-insulator
transition in the normal state, and its relation to
the single particle spectrum.

{\it Acknowledgments:}
We acknowledge use of the High Performance Computing Cluster 
at HRI.  PM acknowledges support from a DAE-SRC 
Outstanding Research Investigator
Award, and the DST India (Athena).


\begin{thebibliography}{99}
\bibitem{bcs-bec-rev} For a recent review, see Q. Chen, J. Stajic, S. Tan
and K. Levin, Phys. Repts. {\bf 412}, 1 (2005).
\bibitem{eagles} D. M Eagles, Phys. Rev. {\bf 186}, 456 (1969).
\bibitem{leggett} A.~J.~Leggett in {\it Modern Trends in the 
Theory of Condensed Matter},  Springer-Verlag, Berlin. 
\bibitem{noz} P.~Nozieres and S. Schmitt-Rink, J.~Low.~Temp.~Phys.
{\bf 59}, 195 (1985).
\bibitem{micnas}
For an early review, see
R. Micnas,  {\it et al.}, Rev. Mod. Phys. {\bf 62}, 113 (1990).
\bibitem{rand-rev} M.~Randeria in {\it Bose-Einstein Condensation},
Cambridge University Press (1995).
\bibitem{qmc-scal1}
R. T. Scalettar, {\it et al.}, Phys. Rev. Lett. {\bf 62}, 1407 (1989).
\bibitem{qmc-moreo1}
A.~Moreo and D.~J.~Scalapino, Phys. Rev. Lett. {\bf 66}, 946 (1991).
\bibitem{qmc-moreo2}
A.~Moreo, {\it et al.}, Phys. Rev. B{\bf 45}, 7544 (1992).
\bibitem{sc-spingap}
M.~Randeria, {\it et al.}, Phys. Rev. Lett.  {\bf 69}, 2001 (1992).
\bibitem{sc-nfl}
N.~Trivedi and M.~Randeria, Phys. Rev. Lett. {\bf 75}, 312 (1995).
\bibitem{t-matrix1} B.~Kyung, {\it et al.},  Phys. Rev. B{\bf 64}, 
075116 (2001).
\bibitem{t-matrix2} H. Tamaki, {\it et al.}, Phys. Rev. A{\bf 77},
063616 (2008).
\bibitem{t-matrix3} J. J. Deisz, {\it et al.}, 
Phys. Rev. B{\bf 66}, 014539 (2002).
\bibitem{dmft1}  M.~Keller, {\it et al.},  
Phys. Rev. Lett. {\bf 86}, 4612 (2001).
\bibitem{dmft2}  M.~Capone, {\it et al.}, 
Phys. Rev. Lett. {\bf 88}, 126403 (2002).
\bibitem{solms} F. Solms, {\it et al.}, 
Phys. Rev. B{\bf 49}, 15945 (1994).
\bibitem{dubi} Y.~Dubi, {\it et al.}, Nature, {\bf 449}, 876 (2007).
\bibitem{dag} M.~Mayr, {\it et al.}, 
Phys. Rev. Lett. {\bf 94}, 217001 (2005).
\bibitem{vnand} V. Singh, {\it et al.}, arXiv:1104.4912.
\bibitem{bdg}
P.~G.~de Gennes, {\it Superconductivity of metals and alloys}, Addison
Wesley (1989).
\bibitem{tca} S.~Kumar and P.~Majumdar, Eur. Phys. J.~B, {\bf 50}, 571 (2006).
\end{thebibliography}
\end{document}